\documentclass{cta-author}

{}
{}
{}

\usepackage[colorlinks = true,
linkcolor = blue,
urlcolor  = blue,
citecolor = blue,
anchorcolor = blue]{hyperref}
\usepackage{graphicx}
\usepackage{subcaption}

\newcommand{\myfig}[1]{Fig. \ref{#1}}

\begin{document}

\supertitle{}

\title{A Machine Learning Based Classification Approach for Power Quality Disturbances Exploiting Higher Order Statistics in the EMD Domain}

\author{\au{Faeza Hafiz~$^{*}$} and \au{Celia Shahnaz~$^{\dagger}$}}

\address{{Department of Electrical and Electronic Engineering,\\ Bangladesh University of Engineering and Technology, Dhaka, Bangladesh}\\
{E-mail: $^{*}$faezahafiz@gmail.com, $^\dagger$celia@eee.buet.ac.bd}
}

\begin{abstract}
The aim of this paper is to propose a new approach for the pattern recognition of power quality (PQ) disturbances based on Empirical mode decomposition (EMD) and $k$ Nearest Neighbor ($k$-NN) classifier. Since EMD decomposes a signal into intrinsic mode functions (IMF) in time-domain with same length of the original signal, it preserves the information that is hidden in Fourier domain or in wavelet coefficients. In this proposed method, power signals are decomposed into IMFs in EMD domain. Due to the presence of non-linearity and noise on the original signal, it is hard to analyze them by second order statistics. Thus, an effective feature set is developed considering higher order statistics (HOS) like variance, skewness, and kurtosis from the decomposed first three IMFs. This feature vector is fed into different classifiers like $k$-NN, probabilistic neural network (PNN), and radial basis function (RBF). Among all the classifiers, $k$-NN showed higher classification accuracy and robustness both in training and testing to detect the PQ disturbance events. Simulation results evaluated that the proposed HOS-EMD based method along with $k$-NN classifier outperformed in terms of classification accuracy and computational efficiency in comparison to the other state-of-art methods both in clean and noisy environment.
\end{abstract}

\maketitle

\section{Introduction}
Reliable power supply is one of the a major concern for smart grids due to the rapid inclusion of sensitive loads into the power system. Degradation of reliability in power system arises because of various reasons, like power-line disturbances, malfunctions, instabilities, short lifetime, failure of electrical equipment, intermittency of distributed energy resources, and so on. All of these reasons can be detected analyzing the voltage/current signals of the power system. For example, faults in a distribution system shows sag or momentary interruption in voltage signals, sudden drop off large load or energization of a large capacitor bank cause swell in voltage, harmonic distortion occurs due to the inclusion of power electronic inverters or solid-state switching devices, capacitor or transformer switching leads to transients, sudden higher load inclusion impacts are seen as flickers, lightning strikes causes spikes in the voltage signal \cite{Mahela2015}, \cite{Mishra2018}. 


In order to ensure an improved power delivery to the customers, utility companies need to know the reasons of disturbances to resolve the issues. Thus it is important for the utility companies to detect the problem from the signals at hand so that they can take mitigation actions within short period of time. In this regard, researchers are interested recently to use efficient and appropriate signal processing methods to extract all the the important information to classify the PQ disturbance events. Most of the proposed research works are based on extracting all the information from a signal into a set of features. Then they utilized different classification methods to obtain the accurate class of a signal based on the feature set. As a result, the identification process of PQ disturbance events follows three steps, signal analysis or decomposition, feature vector selection; and train a classifier to accurately classify the disturbance \cite{Hafiz2012}.

According to the above statement, the available power disturbance time series data is processed through different signal processing approaches in the first step of PQ disturbance classification. Considering this objective, several signal processing transformation methods were proposed in the literature. Among all of them, fourier transform (FT) was most commonly used \cite{Flores}, which is only applicable to the stationary signals. To obtain time frequency information from the PQ disturbance waveform, short time fourier transform (STFT) was proposed in \cite{Salama2001,Bollen,Jurado}. The limitation of STFT is that it can not extract transient signal information properly due to fixed window size selection. Different wavelet based transformations, like wavelet transformation, wavelet packet transformation,  and wavelet multiresolution analysis were analyzed in \cite{Poisson,Bhatt,Santoso,Zhang,Kamaraj,Salama,Gaing}. These proposed wavelet domain based methods are dependent on proper selection of mother wavelet and the level of decomposition for effective recognition of disturbance signals. Stockwel transform was proposed in \cite{Soong,Pani}. It uniquely combines a frequency dependent resolution and simultaneously localizes the real and imaginary spectra. But similar to the STFT, it requires the selection of a suitable window size to match with the specific frequency content of the signal. In \cite{Wen}, the authors described a Hilbert transform based signal decomposition for the feature extraction from the distorted waveform to generate an analytical signal. But Hilbert Transformer provided a better approximate of a quadrature signal only if the signal reached into a narrow band condition. A combination of Prony analysis and Hilbert transform was also considered in \cite{Feliat}, where a signal was reconstructed using linear combination of damped complex exponential. A prediction model was also developed in this work to estimates the different modes of a signal. The estimated signal best fits with the original signal only if condition of minimization of least square error between the original signal and estimated signal was satisfied. One limitation of this method is that the number of mode is required to be selected beforehand for the Prony analysis and no rules were developed to guide the selection procedure of this numbers. Considering these limitations of other methods, we prefer to utilize empirical mode decomposition (EMD) method in this work for PQ disturbance signal analysis. Since EMD is a multi-resolution signal decomposition technique, it has the ability to denoise signals and detect PQ disturbances accurately \cite{Kabir, Shukla2014}. In EMD, the intrinsic oscillatory modes of a signal is identified primarily in time scale. Then the signal is decomposed into intrinsic mode functions (IMFs) according to the oscillatory modes. EMD is adaptive with the basic functions which are derived from the data. The computation of EMD does not require any previously known value of the signal. As a result, EMD is especially applicable for nonlinear and non-stationary signals, such as PQ disturbances. 

\begin{table*}
	\centering
	\caption{Models of power quality disturbance signals}\label{tbl:model_pow_qua_dist}
	\footnotesize
	\begin{tabular}{|l|l|l|}
		\hline
		\textbf{Disturbance} & \textbf{Equations} & \textbf{Parameters}
		\\ \hline \hline
		Normal & $ v(t)=V\sin\omega_{c}t$ & u(t) is the unit function    \\ \hline
		Sag & $ v(t)=V[1-\gamma\{u(t-t_1)-u(t-t_2)\}]\sin\omega_{c}t$ & $0.1\leq\gamma\leq0.9$,
		\\   &        &                                       $T\leq(t_2-t_1)\leq9T$
		\\ \hline
		Swell &  $ v(t)=V[1+\gamma\{u(t-t_1)-u(t-t_2)\}]\sin\omega_{c}t$ & $0.1\leq\gamma\leq0.9$,
		\\   &        &                         $T\leq(t_2-t_1)\leq9T$                          \\   \hline
		Flicker &   $ v(t)=V[1+\gamma\sin(2\pi\beta t)]\sin\omega_{c}t$ & $0.1\leq\gamma\leq0.2$,
		\\   &        &                                      $5\ \text{Hz}\leq\beta\leq 20$ Hz          \\   \hline
		Interruption &  $ v(t)=V[1-\gamma\{u(t-t_1)-u(t-t_2)\}]\sin\omega_{c}t$ & $0.9\leq\gamma\leq1$,
		\\   &        &                         $T\leq(t_2-t_1)\leq9T$                          \\   \hline
		Transient & $v(t)=V[\sin\omega_{c}t +\gamma e^(t-t_1/\tau)\sin\{2\pi f_n(t-t_1)\} \{u(t_2)-u(t_1)\}]$              &  $0.1\leq\gamma\leq0.9$,
		\\  & &  $0.5T\leq (t_2-t_1) \leq 3T$,
		\\   &   &  $300\ \text{Hz}\leq f_n \leq 900\ \text{Hz}$,
		\\    &   &   $8\ \text{ms} \leq \tau \leq 40$ ms            \\    \hline
		Harmonics & $v(t)=V[\sin\omega_{c}t+\gamma_3\sin3\omega_{c}t +\gamma_5\sin 5\omega_{c}t]$ & $0.1\leq\gamma\leq0.9$,
		\\ & & $T\leq(t_2-t_1)\leq9T$,
		\\ & & $0.05\leq\gamma_3,\gamma_5\leq0.15$     \\   \hline
		Sag  & $v(t)=V[1-\gamma\{u(t-t_1)-u(t-t_2)\}]*[\sin\omega_{c}t+\gamma_3\sin3\omega_{c}t +\gamma_5\sin 5\omega_{c}t]$     & $0.1\leq\gamma\leq0.9$,
		\\with  & &   $T\leq(t_2-t_1)\leq9T$,
		\\ Harmonics &  & $0.05\leq\gamma_3,\gamma_5\leq0.15$   \\  \hline
		Swell &  $v(t)=V[1+\gamma\{u(t-t_1)-u(t-t_2)\}]*[\sin\omega_{c}t+\gamma_3\sin3\omega_{c}t +\gamma_5\sin 5\omega_{c}t]$     & $0.1\leq\gamma\leq0.9$,
		\\with &   & $T\leq(t_2-t_1)\leq9T$,
		\\  Harmonics &  & $0.05\leq\gamma_3,\gamma_5\leq0.15$   \\  \hline
		Spike & $v(t)=V[\sin\omega_{c}t-sign(\sin\omega_{c}t)\times\{\sum_{n=0}^{9}\kappa\times\{u(t-(t_1+0.02n))-u(t-(t_2+0.02n))\}\}]$     & $0.1\leq\kappa\leq0.4$,
		\\  & &    $0\leq(t_2,t_1)\leq0.5T$,
		\\ & &   $0.01T\leq(t_2-t_1)\leq0.05T$   \\  \hline
		Notch & $v(t)=V[\sin\omega_{c}t+sign(\sin\omega_{c}t)\times\{\sum_{n=0}^{9}\kappa\times\{u(t-(t_1+0.02n))-u(t-(t_2+0.02n))\}\}]$     & $0.1\leq\kappa\leq0.4$,
		\\  & &    $0\leq(t_2,t_1)\leq0.5T$,
		\\ & &   $0.01T\leq(t_2-t_1)\leq0.05T$   \\  \hline	
	\end{tabular} \label{table:models}
\end{table*}

    
Features selection is the key element among the three steps for PQ disturbance classification. Inappropriate feature selection adds difficulty to the classification. Previous studies overlooked some essential features \cite{Singh,Jayasree,Elango,Chakravorti2017}. In this work, we endeavor to develop appropriate features selection to improve the efficiency of classification. Higher order statistics (HOS) of the extracted IMFs, such as variance, skewness and kurtosis are considered in this regard to form the feature vector. First and second order statistics can not extract the phase information of a signal and can easily be affected by noises, HOS are less effected by background noise and contain phase information. The discriminatory attributes of the HOS for different PQ disturbance signals are more prominent in the EMD domain as seen from the shape of the histograms of the IMFs and the values of the corresponding HOS \cite{Alam,Suri}. Thus, it is expected that HOS of PQ disturbances would be more effective if they are computed in the EMD domain rather than in the time domain. 

Recently reported works applied different machine learning algorithms to classify PQ disturbances after defining the feature vectors from the disturbance waveform. Probabilistic neural network \cite{Bhatt}, radial basis function neural network \cite{Elango}, $k$-nearest neighbour \cite{Pandi}, support vector machines \cite{Lin2018} and decision tree\cite{Achlekar} were mostly utilized classifiers for PQ disturbance signals . In this research, IMFs of the PQ disturbance signals are obtained by using EMD operation. As most frequency content of the PQ disturbance signals lies in the first three IMFs, they are selected for further analysis \cite{Hafiz2013}. HOS of the extracted IMFs, such as variance, skewness and kurtosis are extracted to form the feature vector. The feature set obtained is fed to the radial basis function (RBF), probabilistic neural network (PNN) and $k$ nearest neighbor ($k$-NN) classifiers for classifying the multi class PQ disturbance signals. For the characterization of PQ disturbance signals, mathematical models of eleven classes of disturbances are used. In comparison to the other methods, $k$-NN classifiers shows superior performance for the proposed HOS of EMD (HOS-EMD) based feature vector. Simulation results reveal the effectiveness of the HOS-EMD  method for classifying multi-class PQ disturbance signals .

This paper is organized as follows. The proposed higher order statistics based feature extraction of PQ disturbance signal in EMD domain termed as HOS-EMD method is discussed in Section II. This section includes a brief background of PQ disturbances signal models, how IMFs are generated in EMD domain, feature selection from IMFs, and brief description of $k$-NN, PNN, and RBF classifiers. The simulation results of the HOS-EMD classification method are provided and its performance are compared relative to the other methods in section III. The importance of the proposed method are highlighted in section IV with concluding remarks.

\section{HOS-EMD METHOD}
In power system, a pure voltage or current is a sinusoidal signal that can be mathematically represented as
\begin{equation}\label{Equ:equ1}
v(t)=V\sin\omega_{c}t
\end{equation}

where, $V$ and $f$ represent the amplitude and fundamental frequency respectively.  Different types of power quality disturbance signal like sag, swell, fluctuation, interruption, transient, harmonics, sag with harmonics, swell with harmonics, spike and notch can be seen in power system. The mathematical models of these disturbance signals are provided in Table \ref{table:models}. To classify these signals, proposed HOS-EMD method follows two steps, namely feature extraction and classification. The details of these steps are described below.

\begin{figure}
	\centering
	\includegraphics[width=0.5\textwidth]{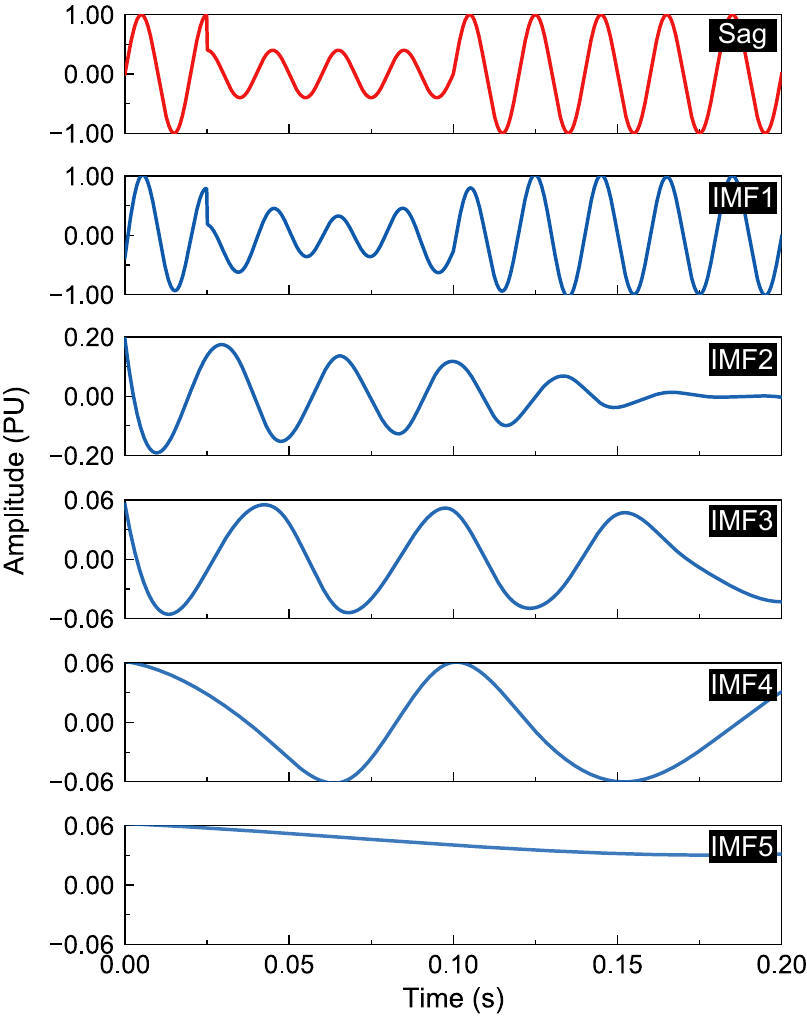}\\
	\caption{\textit{Voltage sag and its intrinsic mode functions}.}
	\label{fig:sag_imf}
    \end{figure}
    
\begin{figure}
	\centering
	\includegraphics[width=0.5\textwidth]{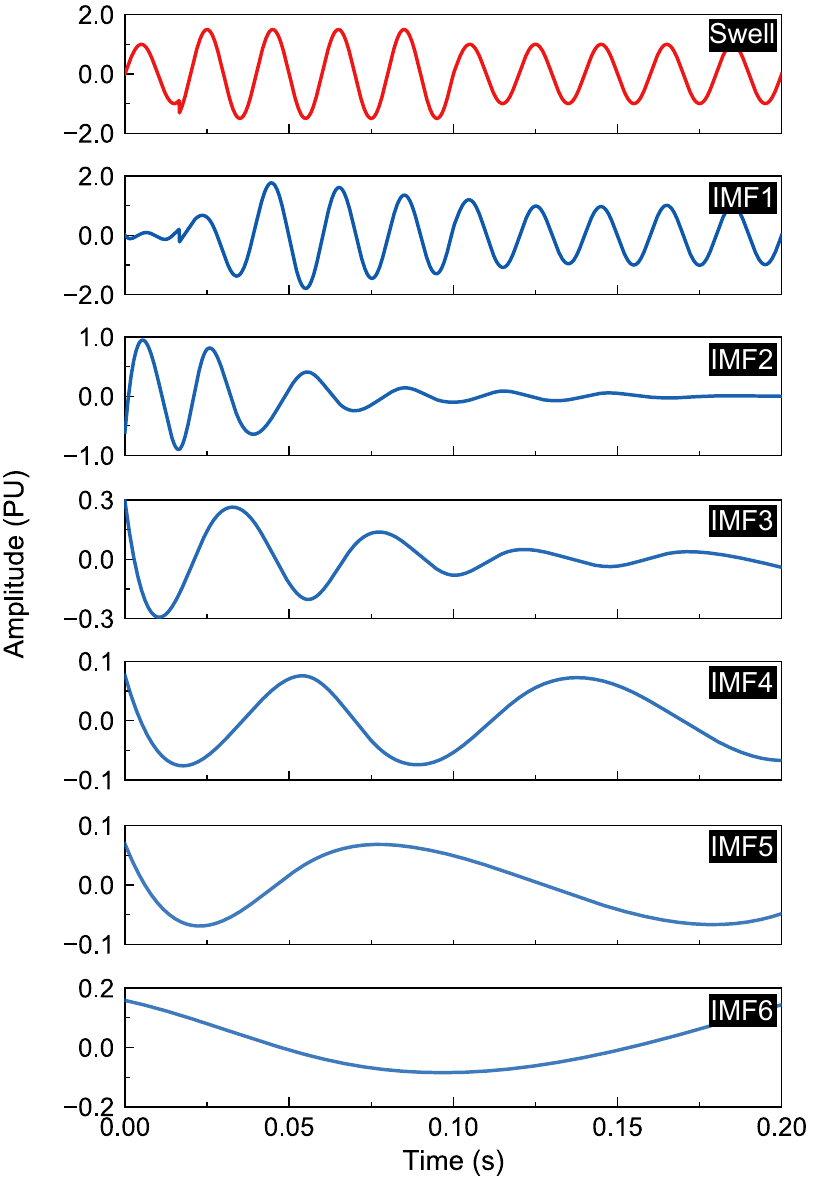}\\
	\caption{\textit{Voltage swell and its intrinsic mode functions}.}
	\label{fig:swell_imf}
    \end{figure}
    
\subsection{Feature Extraction}
\subsubsection{Empirical Mode Decomposition}
EMD is a method that decomposes a signal into multiple IMFs in time domain. As it preserves the domain during decomposition, there is less probability to loose important information from a signal. To decompose a signal into IMFs, the following mandatory conditions are needed to be satisfied. 

$Condition\ 1:$ The total number of local minima and maxima should be equal or differ at most one compared to the number of zero crossings in the whole dataset of the signal. 

$Condition\ 2:$ Mean values of the local minima and local maxima envelop must be zero. 

A brief description of the IMF decomposition is illustrated below.
\begin{enumerate}[i.]
	\item All  the  local  maxima and minima are  determined from the given signal data. 
	\item These data are connected to construct upper and lower envelop of the signal using  by  cubic spline  lines.
	\item The mean of the envelops are calculated as $m_1$. 
	\item The difference between the PQ disturbance signal, $v(t)$ and $m_1$ are calculated as-
	\begin{equation}\label{Equ:equ2}
	e_1(t)=v(t)-m_1
	\end{equation}
	If this difference, $e_1(t)$ satisfies the conditions of IMF described above, then it can be considered as first frequency and amplitude modulated oscillatory mode of $v(t)$.
	\item If  $e_1(t)$ is not  an  IMF,  then it  is passed through the  second  sifting  process,  where  steps  i-iv  are repeated  on  $e_1(t)$ to  obtain another component  $e_2(t)$ by following the equation below:
	\begin{equation}\label{Equ:equ3}
	e_2(t)=e_1(t)-m_2
	\end{equation}
	where $m_2$ is the mean of upper and lower envelopes of $e_1(t)$.
	\item Let after $w$ cycles of operation, $e_w(t)$, is calculated as
	\begin{equation}\label{Equ:equ4}
	e_w(t)=e_{w-1}(t)-m_w
	\end{equation}
	and it satisfies the IMF conditions. The the first IMF component of the original signal will be  $I_1(t)= e_w(t)$.
	\item $I_1(t)$ is then subtracted from $v(t)$, and the residual signal, $r_1(t)$ is calculated as
	\begin{equation}\label{Equ:equ5}
	r_1(t)=v(t)-I_1(t)
	\end{equation}
	This residual signal, $r_1(t)$ is then treated as the original data for calculating the next IMF.
	\item The stopping criteria of the sifting process is calculating the standard different of two consecutive signal and check with some threshold level. Let after repeating steps i-vii for $W$ times, $W$ no. of IMFs is obtained along with the final residue $r_w(t)$ are obtained. Then the standard difference (SD) is calculated as:
	\begin{equation}\label{Equ:equ6}
	SD=\sum^W_{w=2}\frac{|e_{w-1}(t)-e_w(t)|^2}{e_w(t)^2}
	\end{equation}
	where, the index terms, $w$ and $w-1$ are indicating two consecutive sifting processes. Thus the decomposition process is stopped since $r_w(t)$ becomes a monotonic function from which no more IMF can be extracted. To this end, for $W$ level of decomposition, the PQ disturbance signal $v(t)$ can be reconstructed by the following formula,
	\begin{equation}\label{Equ:equ7}
	v(t)=\sum^W_{w=1}I_w(t)+r_w(t)
	\end{equation}
\end{enumerate}

\begin{figure*}
	\centering
	\includegraphics[width=\textwidth]{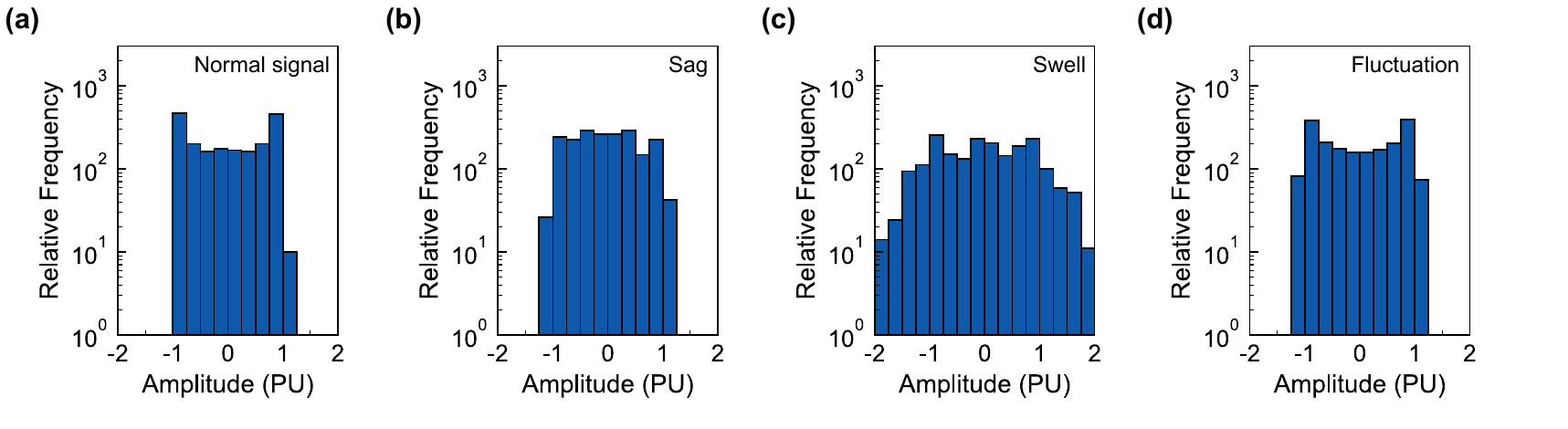}\\
	\caption{\textit{Histogram of first IMF of power quality disturbances. (a) normal signal, (b) sag, (c) swell, and (d) fluctuation}}
	\label{fig:hist}
   \end{figure*}

\subsubsection{IMF Selection}
While decomposing all the PQ disturbance signals through EMD process, sag ang swell always results in six IMFs and harmonics and fluctuation signals are decomposed into one or two IMFs. The IMFs for sag and swell are shown in \myfig{fig:sag_imf} and \myfig{fig:swell_imf}. It is found that frequency contents of a signal are mostly available in the first three IMFs. IMFs become smoother with the increase of their level according to these figures. Thus, we can extract necessary information from first few IMFs instead of considering all of the available IMFs. Eventually, it will help to reduce the requirement of memory and speed up the calculation process for classifiers. Considering this observation, we are motivated to exploit the first three IMFs for feature selection in this work. IMFs are considered as zero for the signals which are decomposed into less than three IMFs. 

\subsubsection{Higher Order Statistics}
\begin{figure}
	\centering
	\includegraphics[width=0.48\textwidth,angle=0]{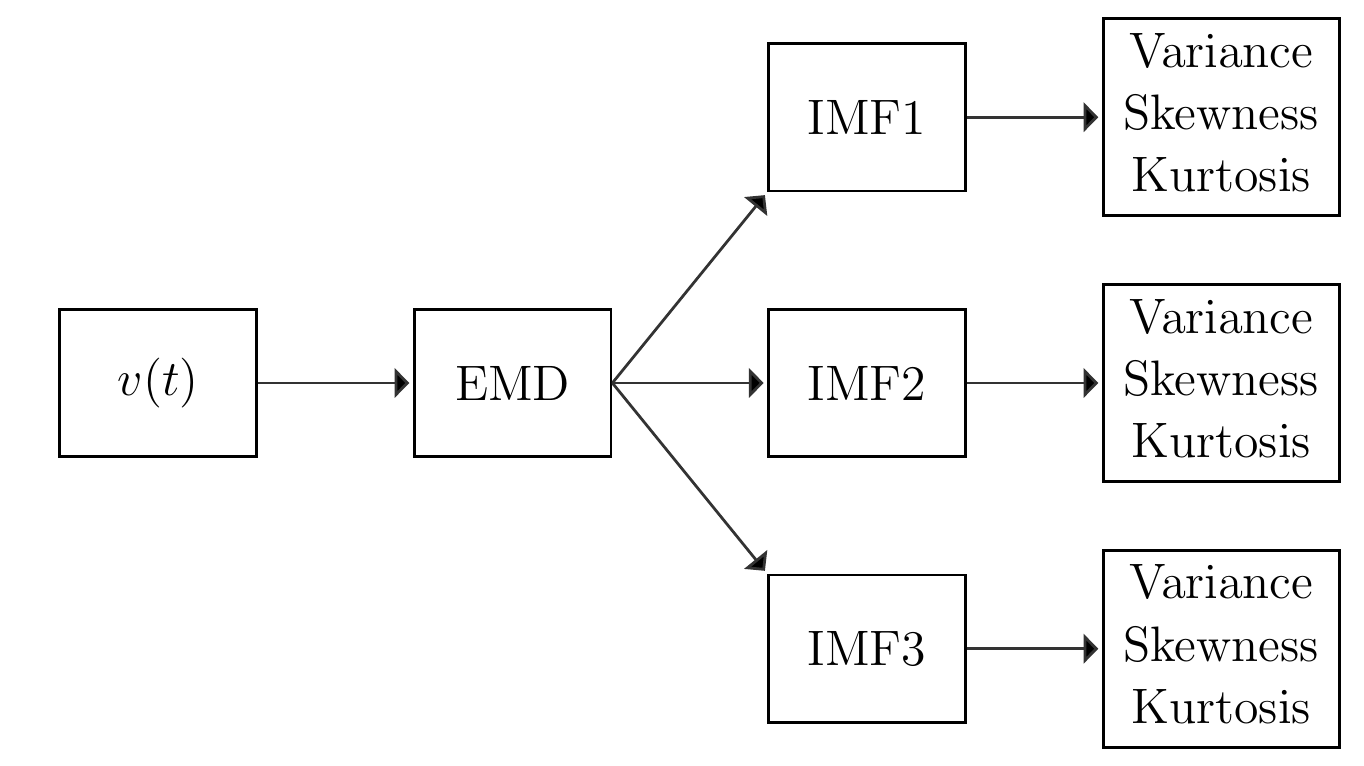}\\
	\caption{\textit{Feature Extraction from distorted waveform}}\label{fig:fig5_EMD_method}
\end{figure}

Distribution of a data set can be understandable easily by analyzing it's level of dispersion, asymmetry and concentration around the mean.Calculation of HOS is an effective way to measure these information, especially for the systems with nonlinear dynamics. They perform better to extract information than the second order statistics with the presence of noise in the signal. In this work, HOS are termed as variance, skewness and kurtosis. From the first three IMFs after decomposition of a PQ disturbance signal in EMD demain, their variance, skewness and kurtosis are calculated as a feature vector for classifying the PQ disturbance signals. For an $N$-point data, ${v_{1}, v_{2},..., v_{N}}$, the corresponding variance ($\sigma^{2}$), skewness ($\alpha_{1}$) and kurtosis ($\alpha_{2}$) are calculated as

\begin{equation}\label{Equ:equ8}
\sigma^{2} =\frac{1}{N}\sum^N_{n=1}(v_{i}-\mu)^{2};\mu =\frac{1}{N}\sum^N_{n=1}(v_{i})
\end{equation}

\begin{equation}\label{Equ:equ9}
\alpha_{1}=\frac{1}{N}\sum^N_{n=1} \left( \frac{v_{i}-\mu}{\sigma} \right)^{3}
\end{equation}

\begin{equation}\label{Equ:equ10}
\alpha_{2}=\frac{1}{N}\sum^N_{n=1} \left( \frac{v_{i}-\mu}{\sigma}\right)^{4}
\end{equation}

here, $\mu$ symbolizes the sample mean of the data. For a symmetric distribution of data about mean, the skewness is zero. Positive skewness occurs when data are spread towards the right of the mean. When data is spread more to the left of the mean, then skewness becomes negative. Kurtosis provides the idea of whether a dataset are heavy-tailed or light-tailed compared to a normal distribution. If distribution of signal is light-tailed, then the kurtosis is less than zero and it becomes greater than zero when the distribution has heavier tails. Figure \ref{fig:hist} shows the histograms of the first IMF of pure signal and three PQ disturbances. From the figures, it is observable that the shapes of the PQ disturbances are different from each other. It is expected that the values of the corresponding variance, skewness and kurtosis are different from each other as these values are delivering the information of the dispersion, asymmetry and tailedness of data. Due to decomposition of the signal into IMFs, HOS based features becomes more prominent in the EMD domain rather than in spatial domain for classifying the PQ disturbance signals. Thus, from the first three extracted IMFs, nine features are derived for the feature vector. Figure \ref{fig:fig5_EMD_method} shows the flow diagram for proposed extracted features from the distorted waveform.

\subsection{Classification}

\subsubsection{$k$-NN Classification}

$k$-NN is a simple and robust classifier \cite{Pandi}. In this algorithm, $k$ neighborhood is defined by the user. For a testing sample, a class is assigned by checking more frequent training samples in the $k$ neighborhood. The value of $k$ is required to be varied to find the match class between training and testing data. The default value of $k$ is 1. In this paper, we varied the value of $k$ from 1 to 10 and the best match is considered. We used euclidean distance to find the object similarity in the $k$ neighborhood as shown in (10).
\begin{equation}\label{Equ:equ12}
\\ d(y_n)=\frac{1}{1+e^{-y_n}}
\end{equation}

\subsubsection{Probabilistic Neural Network}

Similar to the $k$ NN classifier, probabilistic neural networks (PNNs) classifier is a supervised learning classifier. But it follows distinct algorithm which is discussed below \cite{Bhatt}. 

\begin{itemize}
	\item[i.] PNN utilizes the probabilistic model with a Gaussian mapping function.
	\item[ii.] It does not need to set initial weights of the network. But the spread of the Gaussian function is required to be specified.
	\item[iii.] There is no relationship between the learning and recalling processes.
	\item[iv.] Weights of the network do not change with the difference between the inference vector and the target vector.
\end{itemize}

\begin{figure}
	\centering
	\includegraphics[width=0.48\textwidth]{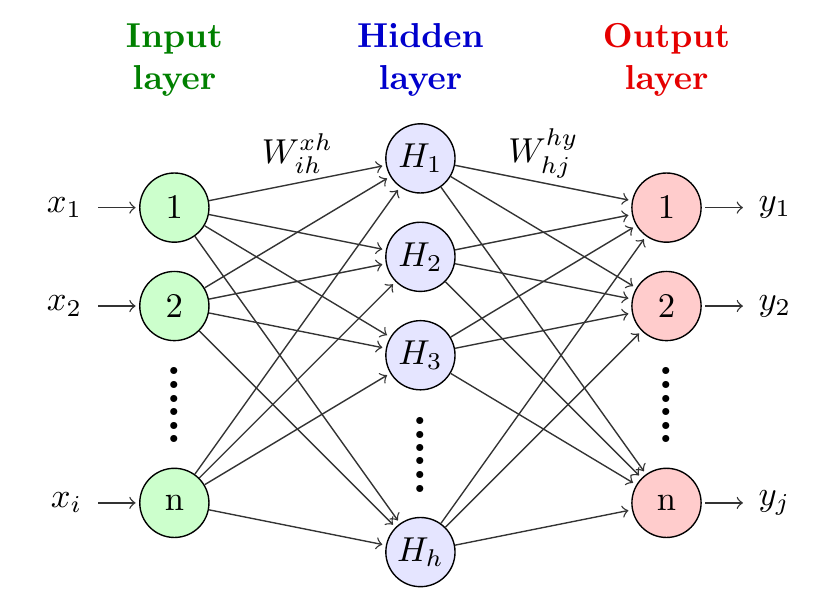}\\
	\caption{\textit{Architecture of PNN neural network}}\label{fig:fig6_archi}
\end{figure}

Figure \ref{fig:fig6_archi} shows architecture of PNN model composed of input, hidden and output layers. For a classification problem, the training data is classified according to their distribution values of probabilistic density function (PDF). A PDF is shown as follows
\begin{equation}\label{Equ:equ12}
f_{r}(x)=\frac{1}{N_{r}}\sum_{j=1}^{N_{r}} {\exp \left(\frac{-(\|X-X_{rj}\|)^{2}}{2\sigma^{2}}\right)}
\end{equation}
Modifying and applying eq. (12) to the output vector $H$ of the hidden layer in the PNN is as
\begin{equation}\label{Equ:equ13}
    H_{h}=\exp \left( \frac{-\sum_{i}(X_{j}-W_{ih}^{xh})^{2}}{2\sigma^{2}} \right)
    \end{equation}
\begin{equation}
net_{j}=\frac{1}{N_{r}}\sum_{h}W_{hj}^{hy}H_{h}\textrm{ and } net_{j}=max_{r}(net_{r})
\end{equation}

\noindent then $y_{j}=1$ or $y_{r}=0$, where

\noindent $i$ = number of input layers;
\newline    $h_j$ = number of hidden layers;
\newline    $j$ = number of output layers;
\newline    $r$ = number of training examples;
\newline    $N$ = number of classifications (clusters);
\newline    $\sigma$ = smoothing parameter (standard deviation);
\newline    $X$ = input vector;
\newline    $\|X-X_{_{rj}}\|$ = Euclidean distance between the vectors $X$ and $X_{rj}$ ; \newline i. e. $\|X-X_{_{rj}}\|$=$\sum_{i}(X-X_{_{rj}})^{2}$
\newline    $W_{ih}^{xh}$= connection weight between the input layer $X$ and and the hidden layer $H$
\newline    $W_{hj}^{hy}$= connection weight between the hidden layer $H$ and the output layer $Y$

\subsubsection{Radial Basis Neural Network }
RBF network consists of an input layer, a hidden layer that uses radial basis functions as activation function, and linear combination of set of weights $W$ in the output layer. The schematic diagram of RBF neural network is shown in \myfig{fig:fig7_RBF}. The transfer functions in the nodes are similar to the multivariate Gaussian density function:
\begin{figure}
	\centering
	\includegraphics[width=0.48\textwidth]{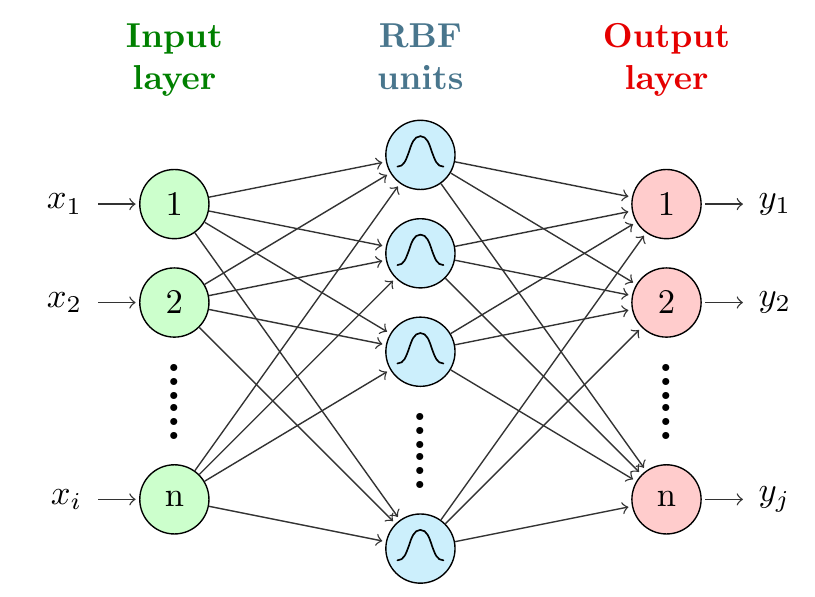}\\
	\caption{\textbf{Architecture of RBF neural network}}\label{fig:fig7_RBF}
\end{figure}

\begin{equation}\label{Equ:equ12}
\Phi_{j}(x) = \exp \left(\frac{\|x-\mu_{j}\|^{^{2}}}{2\sigma_{j}^{2}}\right)
\end{equation}

here $x$ is considered as input vector, $\mu_{j}$ and $\sigma_{j}$ are the center and spread of the multivariate Gaussian function. Each RBF unit has a significant activation function that defines a specific region. This region is determined by $\mu_{j}$ and $\sigma_{j}$. As a result, RBF defines unique local neighborhood in the input space. The connections between the activation layer and the output layer is the linear weighted summation of the RBF units. Thus, the value of $k^{th}$ output node, $y_k$ is followed by given equation:

\begin{equation}\label{Equ:equ12}
y_{k}(x)=\sum_{j=1}^{h}\zeta_{kj}\Phi_{j}(x)+\zeta_{k0}
\end{equation}

here $\zeta_{kj}$ is the connection weight between the $k^{th}$ output and the $j^{th}$ activation layer. $\zeta_{k0}$ is the basis term.

\begin{figure}
	\centering
	\includegraphics[width=0.48\textwidth,angle=0]{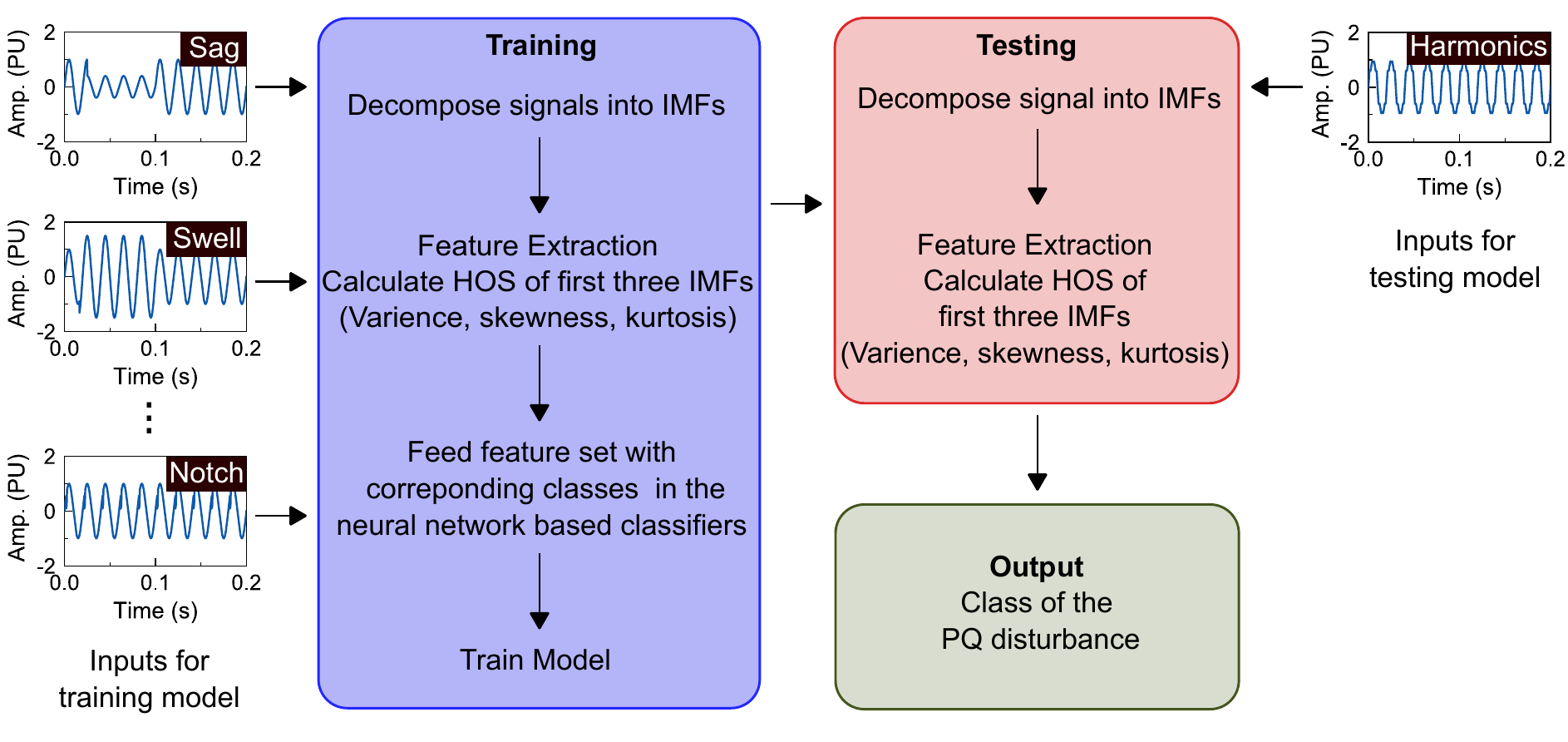}\\
	\caption{\textbf{Overall Method}}\label{fig:fig_overall_method}
\end{figure}

To get the overview of the proposed method, steps discussed above are depicted in \myfig{fig:fig_overall_method}.

\section{SIMULATION RESULT AND ANALYSIS}

For simulation, eleven classes of PQ disturbance signals were generated using in MATLAB considering the equations in Table I with a sampling frequency of 2 kHz.  For convenience, the PQ disturbance signals were termed as:
\begin{enumerate}
	\item~~C1- Normal,
	\item~~C2- Sag,
	\item~~C3 - Swell,
	\item~~C4 - Fluctuation,
	\item~~C5 - Interruption,
	\item~~C6 - Transient,
	\item~~C7 - Harmonics,
	\item~~C7 - Sag with harmonics,
	\item~~C8 - Swell with harmonics,
	\item C10 - Spike,
	\item C11 - Notch.
\end{enumerate}

The effectiveness of the proposed method are described in following subsections.

\subsection{Statistical Analysis of the Feature Extraction}

\begin{figure}
	\centering
	\includegraphics[width=0.49\textwidth,angle=0]{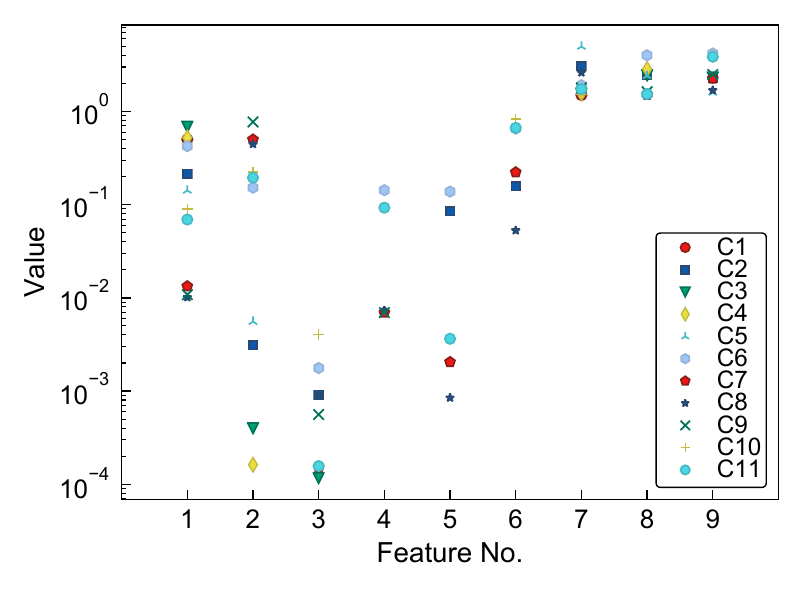}\\
	\caption{\textit{Separability of the HOS-EMD based features in logarithmic scale.}}\label{fig:fig_compactness}
\end{figure}

For the purpose of signal analysis, each PQ disturbance signals are decomposed into IMFs using the algorithm described in subsection 2.1. Then HOS are calculated from the first three IMFs. The impact of decomposition in the EMD domain becomes visible while comparing with the HOS values of the original PQ disturbance signals. Table \ref{table:val_pure_sig} and \myfig{fig:fig_compactness} show the HOS values obtained for different PQ disturbance signals and their first three IMFs, respectively. In \myfig{fig:fig_compactness}, the stamps of the feature set considered in x-axis are variance, skewness, and kurtosis of first three IMFs of different PQ disturbance signals, respectively. For clarification, feature stamp 1, 2 and 3 are the variance, skewness and kurtosis of first IMF. Similarly, variance, skewness, and kurtosis of second IMF are considered as stamp no. 4, 5, and 6;  and third IMF are 7, 8, and 9. Corresponding feature values are shown in logarithmic scale in y-axis. From Table 2, it can be seen that variance, skewness and kurtosis of the original signals for different PQ disturbances are very close to each other. On the other hand, it is clear that the HOS values are distinguishable in EMD domain for different classes of PQ signals from \myfig{fig:fig_compactness}. Since the separability of the PQ disturbances are larger for IMFs than the original signals, the proposed feature set is more robust to differentiate them. It is also clear that with the increase of IMF level, the separability of the features reduces. 

\begin{table}
	\centering
	\caption{Values of Pure Signals}\label{tbl:pure_signal}
	\begin{tabular}{|l|l|l|l|}
		\hline
		\bf{Classes} &  \bf{Variance} & \bf{Skewness} & \bf{Kurtosis} \\ \hline \hline
		C1 & 0.5  &2.77E-16 &0.375 \\ \hline
		C2 & 0.167  &1.11E-18 &0.304 \\ \hline
		C3 & 0.6594  &8.19E-18 &0.6358 \\ \hline
		C4 & 0.5147 & 9.36E-04 &0.7302 \\ \hline
		C5 & 0.45  &1.78E-17 &0.27 \\ \hline
		C6 & 0.5287  &0.152 &0.3586 \\ \hline
		C7 & 0.5  &2.66E-16 &0.4453 \\ \hline
		C8 & 0.3734  &7.78E-18 &0.189 \\ \hline
		C9 & 0.6186  &1.61E-16 &0.6304 \\ \hline
		C10 & 0.5114  &0.0027 &0.3997 \\ \hline
		C11 & 0.4781  &0.0059 &0.3391 \\ \hline
	\end{tabular}
	\label{table:val_pure_sig}
\end{table}

\subsection{Comparative Analysis with Other Methods}

Machine learning based classifiers are required to be trained with large amount of data before testing. Due to unavailability of required amount of database for different classes of PQ disturbance, the classifiers are trained with synthetic data which are generated using the mathematical models along with the random variation of parameters like amplitude ($\gamma$), frequency($\beta$), duration($t_1$,$t_2$) which are described in Table 1. To evaluate the performance of the proposed HOS-EMD method for classification of eleven PQ events, a total of 1485 signals with 135 signals of each class are generated. The fundamental frequency of the signals is 50 Hz.Among 135 signals for each class, 35 signals are utilized for training and the rest of the signals are considered for testing and validation. The performance evaluation are done based on confusion matrix and overall efficiency in percentage (\%) calculation. Confusion matrix is a form of representing the result from a classification exercise. Overall efficiency is calculated using the formula given as follows-

\begin{equation}\label{Equ:equ12}
\textrm{Overall Efficiency}=\frac{\textrm{No. of correctly classified events}}{\textrm{Total no. of events}}
\end{equation}

\begin{figure*}[!t]
	\centering
	\includegraphics[width=\textwidth]{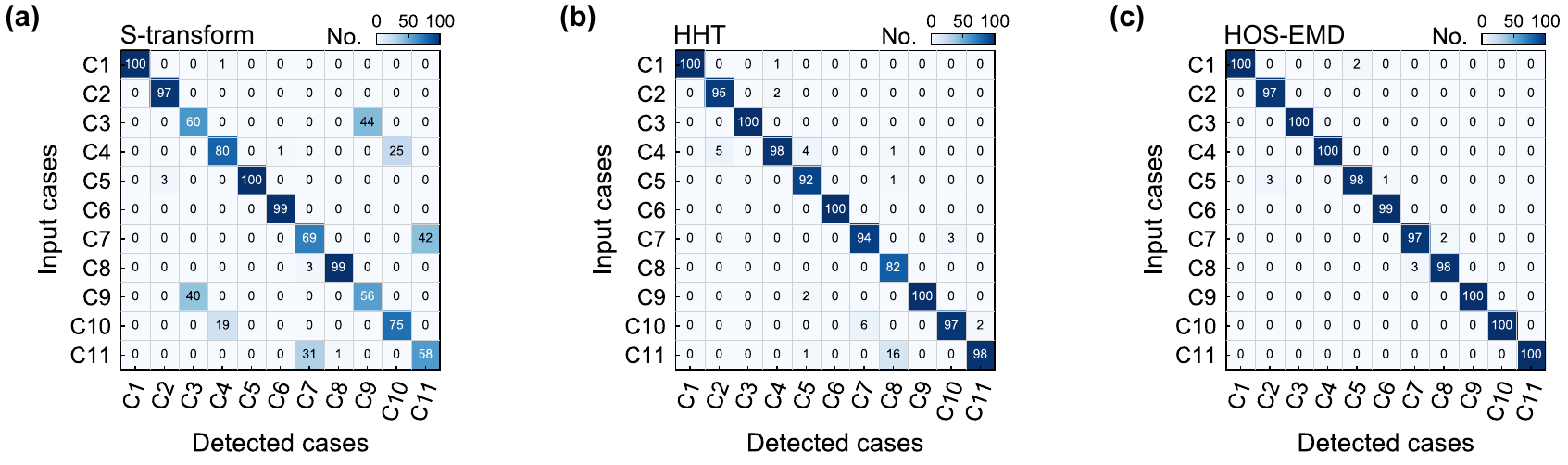}\\
	\caption{\textit{Confusion matrix of different methods: (a) $S$-transform, (b) HHT, and (c) HOS-EMD.}}
	\label{fig:fig9_conf_matrix_part1}
\end{figure*}

For same type of training and testing data a comparative study between $S$-transform \cite{Demir}, Hilbert-Huang transform (HHT) \cite{Manjula}, and proposed HOS-EMD is made. Confusion matrix resulting from the proposed feature set and compared methods set via $k$-NN classifier are presented in \myfig{fig:fig9_conf_matrix_part1}. 

\begin{figure*}[!t]
	\centering
	\includegraphics[width=\textwidth]{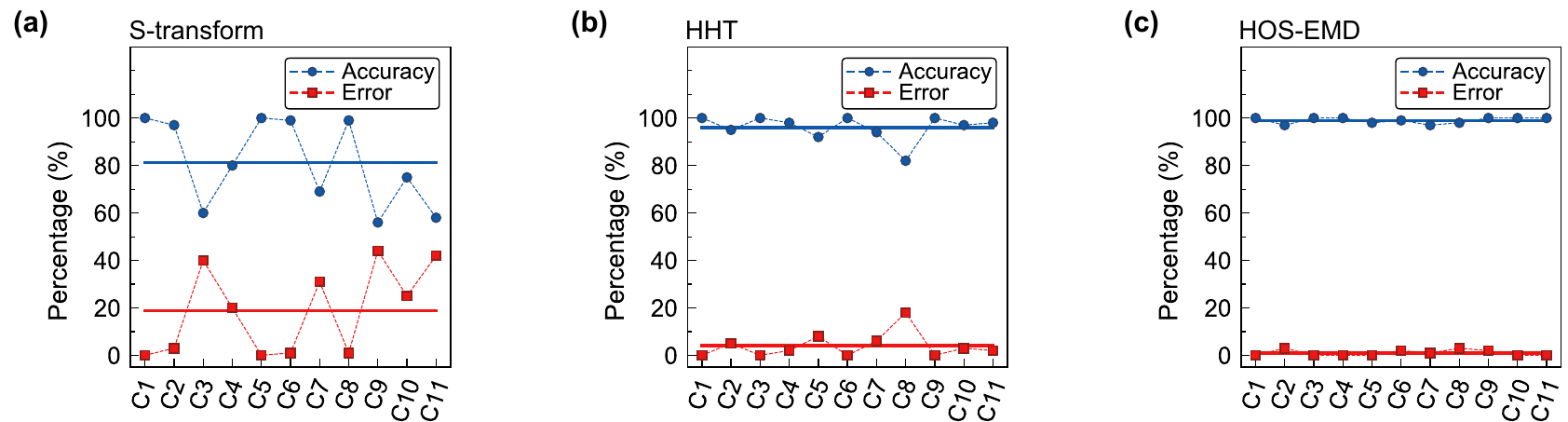}\\
	\caption{\textit{accuracy and error of different methods: (a) $S$-transform, (b) HHT, and (c)HOS-EMD.}}
	\label{fig:fig10_conf_matrix_part2}
\end{figure*}

It can be seen from the diagonal entries of confusion matrix in \myfig{fig:fig9_conf_matrix_part1} (a) that $S$-transform is unable to distinguish among PQ disturbance signals, such as swell (C3), fluctuation (C4), harmonics (C7), swell with harmonics (C9), spike (C10) and notch (C11). Overall accuracy for $S$-transform is 81.2\% which is shown in \myfig{fig:fig10_conf_matrix_part2} (a).  It is also notable from \myfig{fig:fig9_conf_matrix_part1} (b) that HHT based $k$-NN classification misclassifies some sag (C2), interruption (C5), harmonics (C7) and sag with harmonics (C8) signals. Compared to the $S$-transform, HHT improves the classification accuracy up to 96\%. The accuracy level for HHT is depicted in \myfig{fig:fig10_conf_matrix_part2} (b). \myfig{fig:fig9_conf_matrix_part1} (c)  shows that HOS-EMD method based features are able to identify all of the signals almost perfectly. Overall accuracy level is reached up to 99\%. According to \myfig{fig:fig10_conf_matrix_part2} (c), error level is reduced for the proposed HOS-EMD method compared to the $S$-transform and HHT method. 

\begin{figure*}[!t]
	\centering
	\includegraphics[width=\textwidth]{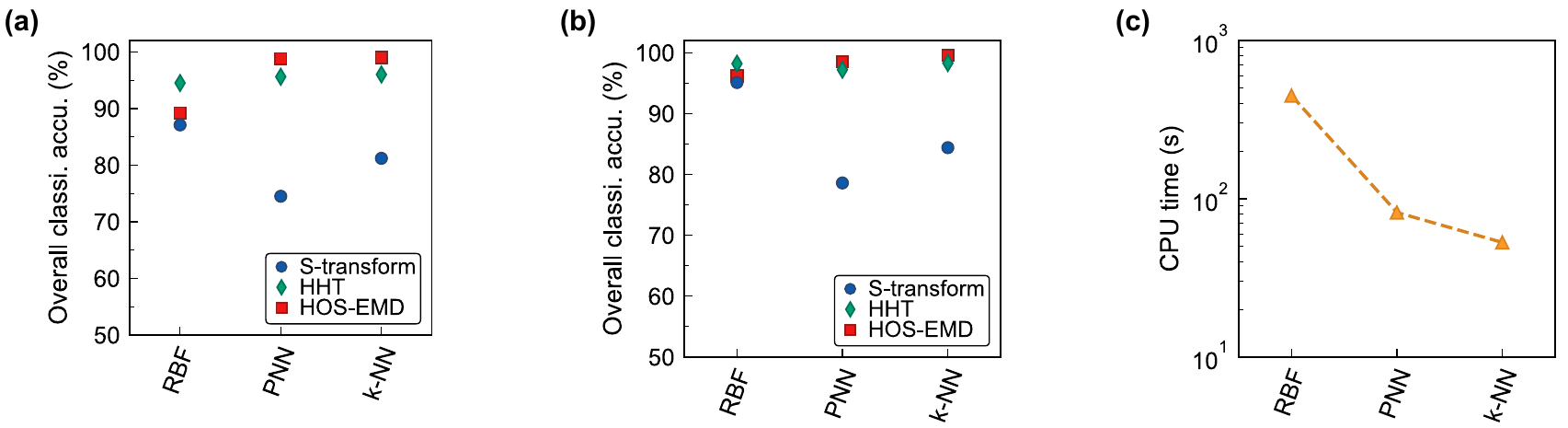}\\
	\caption{\textit{Classification performance for RBF, PNN and $\mathbf{k}$-NN: (a) accuracy when 35 data in training and 100 data for testing are considered, (b) accuracy when 70 data in training and 200 data for testing are considered, and  (c) CPU time requirement for (a).}}
	\label{fig:fig9_classi_res}
\end{figure*}

\subsection{Classifiers Efficiency Analysis}
Classification performance in terms of overall efficiency (\%) when fed to RBF, PNN and $k$-NN classifiers are calculated for all classes. They are presented in \myfig{fig:fig9_classi_res} (a). It is visible that for the selected three classification methods, $S$-transform performs poorly compared to HHT and proposed HOS-EMD method. For RBF classifiers, HHT performed better compared to $S$-transform and HOS-EMD method. But for PNN and $k$-NN classifiers, HOS-EMD method outperformed compared to HHT and $S$-transform. Overall classification accuracy increases for PNN and $k$-NN classifiers compared to RBF during the application of HHT and HOS-EMD. Simulation analysis was also performed by increasing the training and testing data (70 data for training and 200 data for testing). Due to the increase of training data, classification accuracy increased for all of the methods and classifiers according to \myfig{fig:fig9_classi_res} (b). Similar to the previous scenario, $S$-transform performs poorly compared to the other methods. Though HHT performed better while utilizing RBF, overall classification accuracy increased up to 99.6\% while HOS-EMD and $k$-NN classifier is used.

It should be noted that the structure of $k$-NN is simple and it requires less learning time requirement compared to PNN and RBF. The time requirement for training and testing are also specified in \myfig{fig:fig9_classi_res} (c). It is clarified that with the nine features resulting from HOS-EMD domain with $k$-NN classifier requires less time for computation compared to RBF and PNN classifiers. Overall, $k$-NN classifier effectively classifies different kinds of PQ disturbances.

\subsection{Performance of $k$-NN under Noisy Environment}

\begin{figure}[!t]
	\centering
	\includegraphics[width=0.33\textwidth]{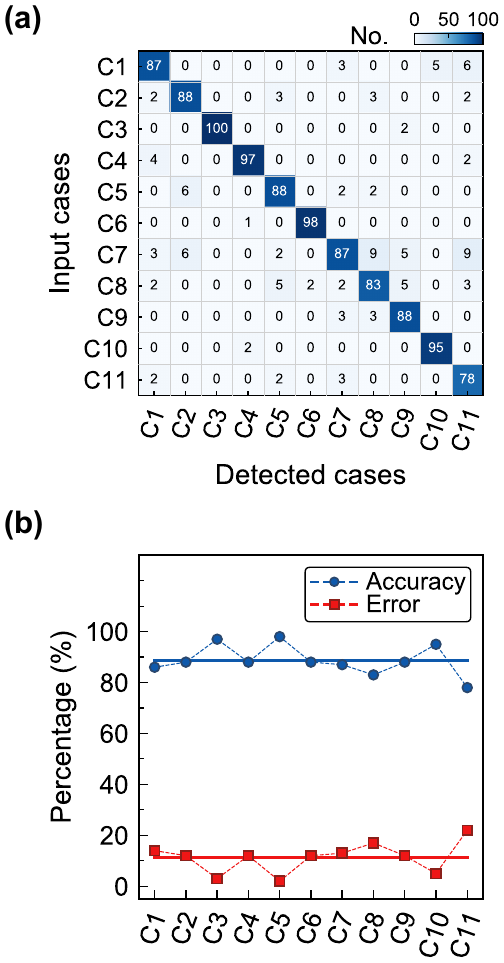}\\
	\caption{\textit{ Performance after inclusion of noise. (a) Confusion matrix, and (b) accuracy and error rate.}
	\label{fig:fig10_conf_err_noise}}
\end{figure}

\begin{figure}
	\centering
	\includegraphics[width=0.31\textwidth]{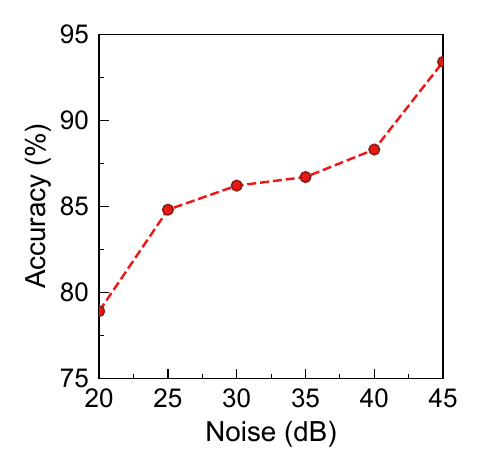}\\
	\caption{\textit{Impact of SNR on classification accuracy.}}\label{fig:fig8_Noise_Level}
\end{figure}

In an electrical power distribution network, the practical data consists of noise. Therefore, the proposed approach has to be analyzed under noisy environment. Gaussian noise is widely considered in the research for power quality issues  \cite{Kabir, Shukla2014}. Thus simulation is also performed with the addition of noise with pure signals and analyzed with EMD-transform for the HOS based feature extraction. $k$-NN is trained and subsequently tested for classification after the extraction of the features. In \myfig{fig:fig10_conf_err_noise} (a), confusion matrix of the proposed HOS-EMD method with $k$-NN classifier is shown which is performed with the inclusion of signal to noise ratio (SNR) of 40 dB. It is notable that classification accuracy decreases and error increases due to the inclusion of noise. Overall accuracy is decreased from 99\% to 93\% according to \myfig{fig:fig10_conf_err_noise} (b). Simulation is also performed with the inclusion of 25, 30, 35 and 45 dB of SNR. The overall classification accuracy results for the HOS-EMD along with $k$-NN classifier are provided in \myfig{fig:fig8_Noise_Level}. It shows that the accuracy level decreases with the increase of noise level. But classification results of our proposed method are quite satisfactory till the inclusion of 25dB of SNR.

\section{CONCLUSION}

The novel contribution of this paper is showing the impact of HOS based features extraction from EMD domain and using $k$-NN classifier to classify PQ disturbance signals. Only the first three IMFs are considered to derive the outcomes from which HOS termed as varience, skewness and kurtosis are calculated to form an effective feature set. The work here is formulated for eleven class problem. The proposed method is compared to the other methods using $S$-transform and Hilbert Huang transform along with PNN and RBF classifiers. It is found that the proposed method shows superior performance in classifying different PQ disturbance signals. The robustness of the proposed method is also verified for noisy environment. For further analysis, the proposed method can be employed for online PQ-disturbance classification.

    \bibliographystyle{iet}

\end{document}